\documentclass[12pt,a4paper]{elsarticle1}

\biboptions{sort&compress}

\usepackage{amssymb}
\usepackage{amsmath}
\usepackage{subfig}
\usepackage{graphicx}
\usepackage{color}

\newcommand{\subimage}{0.5}

\begin{document}
\title{Equilibrium Pricing in an Order Book Environment:\\
       Case Study for a Spin Model}

\author[ude]{Frederik Meudt}

\author[ude]{Thilo A.~Schmitt}

\author[ude]{Rudi Sch\"afer\corref{cor1}}

\author[ude]{Thomas Guhr}

\cortext[cor1]{corresponding author, e--mail: rudi.schaefer@uni-due.de}

\address[ude]{Fakult\"at f\"ur Physik, Universit\"at Duisburg--Essen, Duisburg, Germany}

\date{\today}

\begin{abstract}
  When modelling stock market dynamics, the price formation is often
  based on an equilbrium mechanism. In real stock exchanges, however,
  the price formation is goverend by the order book. It is thus
  interesting to check if the resulting stylized facts of a model with
  equilibrium pricing change, remain the same or, more generally, are
  compatible with the order book environment. We tackle this issue in
  the framework of a case study by embedding the
  Bornholdt--Kaizoji--Fujiwara spin model into the order book
  dynamics. To this end, we use a recently developed agent based model
  that realistically incorporates the order book.  We find realistic
  stylized facts. We conclude for the studied case that equilibrium
  pricing is not needed and that the corresponding assumption of a
  ``fundamental'' price may be abandoned.
\end{abstract}


\maketitle

\section{Introduction}
\label{sec1}

In models for financial markets, the traders' decision to buy or to
sell is typically mapped to supply and demand, which are then balanced to determine the resulting price change,
see~\cite{de1990noise,caldarelli1997prototype,lux1999scaling,cont2000herd}.
This equilibrium pricing is a classical
concept in the economics literature~\cite{fama1970efficient,Mas-Colell1995}. On the
other hand, a real stock exchange uses a double auction order book. It
provides a trading platform to every registered participant in which
all offers to buy or sell are listed, ensuring that all traders have
the same information.  A price is quoted whenever a buy or sell orders
match. Thus, the real price formation is quite different from the
concept of equilibrium pricing.  Here, we confront a model based on
equilibrium pricing with the dynamics of a double auction order
book. We wish to check the model results in view of the more realistic
price formation.

Due to the rich variety of existing equilibrium pricing models, we
have to restrict ourselves to case studies. In a previous
work~\cite{Wagner2014347}, we looked at a rather simple--minded decision
making model that we analyzed with equilibrium pricing and,
alternatively, in an order book environment. Here, we wish to address
the more advanced Ising--type--of spin models which are known to
properly capture several aspects of financial markets.  We choose the
Bornholdt--Kaizoji--Fujiwara
model~\cite{doi:10.1142/S0129183101001845,Kaizoji2002441} as a
particularly interesting representative. We sketch its salient features
in Sec.~\ref{sec2}. To apply order book dynamics to this model, we
employ an agent--based model that fully accounts for the double
auction order book as used in stock exchanges.  It obviously makes
sense to choose a ``minimalistic'' agent--based model free of
additional features that might influence the resulting picture.  Such
an agent--based model was recently put forward and successfully tested
in Ref.~ \cite{0295-5075-100-3-38005}. In Sec.~\ref{sec3}, we briefly
present its setup and adjust it to the Bornholdt--Kaizoji--Fujiwara
model. This amounts to applying the decision making part of the
Bornholdt--Kaizoji--Fujiwara model, but to then let the order book
work. In particular, this lifts the constraints due to equilibrium
pricing. We statistically analyze selected resulting
quantities in Sec.~\ref{sec4}. We summarize and conclude in
Sec.~\ref{sec5}.

\section{Bornholdt--Kaizoji--Fujiwara Model}
\label{sec2}

Considering their paramount success for the study of phase transitions
in statistical mechanics, it is not surprising that the application of
spin models, in particular those of the Ising type, to financial
markets has a long history,
see~\cite{doi:10.1142/S0129183101001845,Kaizoji2002441,doi:10.1142/S0129183199000930,Chowdhury1999,Roehner2000,Kaizoji2000493,doi:10.1142/S0129183102003000,doi:10.1142/S0129183102002961}.
Bornholdt~\cite{doi:10.1142/S0129183101001845} introduced an
additional coupling constant to the ferromagnetic nearest neighbor
interaction which then couples the individual spins to the total
magnetization. This is motivated by two conflicting economic
driving forces:
\begin{enumerate}
\item ``Do what your neighbors do'', by aligning your spin to your neighbors.
\item ``Do what the minority does'', by coupling to the magnetization.
\end{enumerate}
As there is no quantifiable stock price in the original version of the
model~\cite{doi:10.1142/S0129183101001845}, Kaizoji, Bornholdt and
Fujiwara~\cite{Kaizoji2002441} extended it accordingly by setting up a
stock market with two groups of traders. We refer to this version 
as Bornholdt--Kaizoji--Fujiwara model.

There are $n$ \textit{interacting traders} $i$ whose investment
attitude is represented by one spin variable $S_i(t)=\pm 1, \
i=1,\ldots,n$ each. The dynamics of the spins is governed by a heat
bath that depends on a local Hamiltonian $h_i(t), \ i=1,\ldots,n$ 
and determines a probability $q$ such that
\begin{eqnarray}
S_i(t+1) & = +1 & \quad \text{with} \qquad q=\frac{1}{1+\exp{(-2\beta h_i(t))}}\\
S_i(t+1) & = -1 & \quad \text{with} \qquad 1-q \ .
\label{heatbath}
\end{eqnarray}
Here, $S_i(t)=1$ ($S_i(t)=-1$) represents a positive (negative)
investment attitude, meaning that trader $i$ buys (sells) the stock.
The traders' perception of the market is driven by two kinds of
information. Locally, he is only influenced by the nearest interacting
traders, but globally, it will also affect him whether or not he
belongs to the majority group. This is measured by the absolute value
$|M(t)|$ of the magnetization
\begin{equation}
M(t)=\frac{1}{n}\sum^{n}_{i=1}S_i(t) \ .
\end{equation}
To accumulate wealth, the trader has to be in the majority group and
the majority has to expand over the next trading period. However, if
$|M(t)|$ already has a large value, further increase is
hampered. Traders in the majority group then tend to switch to the
minority to avert a loss. On the other hand, a trader in the minority
group tends to switch to the majority in the quest for
profit. Altogether, the larger $|M(t)|$, the larger is the tendency
for any group member to switch sides. The local Hamiltonian $h_i(t)$
entering Eq.~(\ref{heatbath}) reads 
\begin{equation}
h_i(t)=\sum^{n}_{j=1}J_{ij}S_j(t)-\alpha S_j(t)|M(t)| \ ,
\end{equation}
with interaction $J_{ij}=J$ for nearest neighbors $i,j$ and with
$J_{ij}=0$ for all other pairs $i,j$. The global coupling constant $\alpha$
is positive, $\alpha>0$.

How is the stock price $p(t)$ determined in this model? --- A number
$m$ of \textit{fundamentalist traders} is introduced whose decisions
are driven by supply and demand. They assume to have a reasonable
knowledge of the fundamental value $p^{*}(t)$ of the stock price.  If
the price $p(t)$ falls below that threshold $p^{*}(t)$, the
\textit{fundamentalists} buy the stock, otherwise they sells it. The
\textit{fundamentalists'} excess demand is given by
\begin{equation}
x^F(t) = a m \Big(\log p^*(t) - \log p(t)\Big) 
       = a m \log\frac{p^*(t)}{p(t)} \ ,
\end{equation}
where $a$ is a parameter characterizing how strongly the
fundamentalists react to the price difference between fundamental
value and current price.  On the other hand, the \textit{interacting
  traders'} excess demand is governed by the total magnetization,
\begin{equation}
 x^I(t)=bnM(t) 
\end{equation}
with $b$ being the corresponding strength parameter. 

The crucial assumption in the Bornholdt--Kaizoji--Fujiwara model
is now the balance of supply and demand such that
\begin{align}
 0&=x^F(t)+x^I(t)\\
  &= am \log\frac{p^*(t)}{p(t)} + bnM(t) \ .
\end{align}
From this the relative price change $r_l(t)$ after the time step from $t$
to $t+\Delta t$ follows according to
\begin{align}
r_l(t)&= \log\frac{p(t+\Delta t)}{p(t)} = \log p(t+\Delta t) -\log p(t)\\
&=\log\frac{p^*(t+\Delta t)}{p^*(t)} + \lambda \Big(M(t+\Delta t)-M(t)\Big) 
\end{align}
with the combination of constants as given by
\begin{equation}
\lambda = \frac{bn}{am} \ .
\end{equation}
Here, we slightly differ from Ref.~\cite{Kaizoji2002441} where the
price change is defined from the time step $t-\Delta t$ to $t$ with
fixed $\Delta t=1$.  For the sake of simplicity, we set $p^*(t)$
constant implying
\begin{equation}
 r_l(t)=\lambda \Big(M(t+\Delta t)-M(t)\Big) \ .
\end{equation}
As the above sketch shows, the price $p(t)$ in the
Bornholdt--Kaizoji--Fujiwara model results from a supply and demand
mechnism. The authors establish a link bewteen magnetization and
trading volume and give an interpretation of the aperiodic switching
between bull and bear markets. They also show in a detailed analysis
that their model reproduces, in an impressive fashion, stylized facts
of real financial data, see Ref.~\cite{doi:10.1080/713665670}, such as
clustered volatilities, positive cross--correlation between trading
volume and volatility, powerlaw fat tails and certain similarities of
the volatilities at different time scales.

Nevertheless, the concept of a fundamental price $p^{*}(t)$, which
may be interpreted in the spirit of the ``fair price'' in the efficient
market model~\cite{fama1970efficient}, raises questions. Why should such a fair
or fundamental price exist at all? --- It is an artificial criterion
applied from outside the market. Is this consistent with the reality
in which ``the market makes the price''? --- The zero intelligence
trading model~\cite{Farmer08022005,farmer2006random} works without such an external concept and
still produces results which are equivalent to those of the efficient market model. Consequently, it is interesting to
study the Bornholdt--Kaizoji--Fujiwara model in a realistic setting by
dropping the concept of the fundamental price and applying the full
order book dynamics instead. Put differently, we only use the decision
making procedure of the Bornholdt--Kaizoji--Fujiwara model and leave
the rest to the trading via the order book.

\section{Agent--Based Model}
\label{sec3}

After rewieving the model setup of Ref.~\cite{0295-5075-100-3-38005}
in Sec.~3.1, we introduce in Sec.~3.2 the IsingTrader that is
especially adjusted to the Bornholdt--Kaizoji--Fujiwara model.

\subsection{Sketch of the Model}
\label{sec31}

Agent--based models provide a useful microscopic framework for the
understanding of stylized facts, even though their complexity often
outrules a one--to--one assignment of input and output.  There are
numerous agent--based models for financial markets involving different
approaches, see
\textit{e.g.}~\cite{doi:10.1080/713665670,doi:10.3905/jpm.1989.409233,Levy1994103,Lebaron02buildingthe,PhysRevE.76.016108}.

In Ref.~\cite{0295-5075-100-3-38005}, a ``minimalistic'' agent--based
model was introduced that implements a double--auction order book by
only resting upon absolutely essential features. It posseses a variety
of traders that follow a given set of rules.  The order book stores
the limit orders ascending from the cheapest buy to the most expensive
sell order. Prices and time are discretized to the tick--size and 
simulation steps, respectively. Orders are cleared,
whenever they are marketable, \textit{i.e.}, whenever some buy and
sell orders match. In each simulation step an arbitrary number of
traders are active, the order of the trading actions is
randomized. Each active trader can place one order and draws his next
time of action or rather his waiting time $t_{\textrm{wt}}$ from an
exponential distribution
 \begin{equation}
  f(t_{\textrm{wt}})=\frac{1}{\mu_{\textrm{wt}}}
            \exp(-t_{\textrm{wt}}/\mu_{\textrm{wt}}) 
            \qquad \text{with} \qquad \mu_{\textrm{wt}}=cN
\label{eq:waitingtime}
\end{equation}
at the end of his trading action. The mean value $\mu_{\textrm{wt}}$
is the product of the number $N$ of traders and a parameter $c$
calibrated to achieve approximately 5.4 trades per minute. This choice
coincides with the average trade frequency of the top $75\%$ stocks in
the S\&P 500 index traded in 2007~\cite{0295-5075-100-3-38005}. (To
avoid confusion, we mention that the distributions denoted $f$ in the
present study are denoted $p$ in Ref.~\cite{0295-5075-100-3-38005}.)
Limit orders are placed with a lifetime $t_{\textrm{lt}}$ drawn from
an exponential distribution
\begin{equation}
 f(t_{\textrm{lt}})=\frac{1}{\mu_{\textrm{lt}}}\exp(-t_{\textrm{lt}}/\mu_{\textrm{lt}}) 
\label{eq:lifetime}
\end{equation}
with mean value $\mu_{\textrm{lt}}$. We notice that traders of
different type can choose different lifetimes $t_{\textrm{lt}}$
corresponding to their governing rules.  The virtual trading is
carried out by the RandomTrader whose buy or sell limit orders are
random with normal distributed prices centered around the current best
price. The order sizes $\nu$ are exponentially distributed,
\begin{equation}
 f(\nu)=\frac{1}{\mu_{\textrm{vol}}}\exp(-\nu / \mu_{\textrm{vol}})
\label{eq:bestprice}
\end{equation}
with mean value $\mu_{\textrm{vol}}$. Short--selling is allowed and
the traders have unlimited credit.

In Ref.~\cite{0295-5075-100-3-38005}, a gap structure in the order
book is identified as the reason for fat tails. Two mechanisms yield
such gaps: canceling of older limit orders and orders placed far away
from the current midpoint. The more liquidity is provided by limit
orders, the less likely are extreme price shifts. The finiteness of
lifetimes $t_{\textrm{lt}}$ of limit orders placed by the
RandomTraders produces gaps, if the lifetime $t_{\textrm{lt}}$ is in
the range of the rate at which new orders are placed. Consistent with
Farmer~\textit{et al.}~\cite{doi:10.1080/14697680400008627}, price
gaps between limit orders are at low liquidity even relevant close to
the current midpoint. It also happens that large order volumes yield
fat tails because orders far away from the midpoint are very unlikely
and lead to large gaps, implying that the liquidity deep in the order
book is very low. These gaps can only be reached with very large
volumes which explains the observations.  For very small order
lifetimes $t_{\textrm{lt}}$, the number of limit orders becomes so
small that the order book effectively plays a minor role. The price
formation is then mainly driven by the specific behavior of the
trader, in particular by the distribution used to determine the order
price. To study the relevance of the order book, we should not focus
on this regime.  Following Ref.~\cite{0295-5075-100-3-38005}, we only
consider order lifetimes which are sufficiently large, \textit{i.e.},
$\mu_{\textrm{lt}} >40$ time steps.

\subsection{IsingTrader}
\label{sec32}

We now introduce the IsingTrader as a new trader type in the
agent--based model of Ref.~\cite{0295-5075-100-3-38005}. With the
IsingTrader we implement the Bornholdt--Kaizoji--Fujiwara
model apart from the concept of the fundamental price.
The order book dynamics alone generates the price.

We identify every lattice site in the Bornholdt--Kaizoji--Fujiwara
model as an independent IsingTrader.  The spin $S_i(t)$ of every
IsingTrader determines, when he is active, whether he will buy or sell
in the particular time step. We do not modify the mechanism that
activates the traders to avoid interference with the time evolution in
the agent--based model, and neither do we change the drawing of the
order volumes as compared with the RandomTrader, because the
Bornholdt--Kaizoji--Fujiwara model does not provide a corresponding
appropriate rule.  We also have to decide which typ of order the
IsingTrader places. According to the Bornholdt--Kaizoji--Fujiwara
model, we ought to derive an appropriate limit price and a lifetime to
place limit orders. This, however, would be quite complicated and,
importantly, not in line with our intention to abandon the concept of
equilibrium pricing. The better choice is the market order, because the
trader's spin value in the Bornholdt--Kaizoji--Fujiwara model directly
translates to a demand at the current point in time. This means in the
framework of the agent--based model that the trader has to buy or sell
the stock directly, depending on his spin value. This can only be
achieved by a market order. Another question arises regarding the
lattice dynamics. In the agent--based model, we wish to affect the
lattice dynamics as little as possible. We have to account for the
facts that, first, not every IsingTrader is active in every timestep
and that, second, too many market orders at one time would wipe the
whole order book clean. Hence, it is reasonable to couple the rate of
sweeps over the spin grid to the trading dynamics in the agent--based
model by a sweep probability $q_{\text{sweep}}$.

For comparative reasons, we also introduce the
LiquidityTaker~\cite{Schmitt2011} who is a RandomTrader placing market
orders. Comparing the results for IsingTrader and LiquidityTaker, we
are able to trace statistical features back to the way how the
IsingTrader makes his decisions.

\section{Results}
\label{sec4}

We performed simulations with $N_\textrm{Random}=2160$ RandomTraders
with limit-order lifetimes of $\mu_{\textrm{lt}}=600$ time steps,
because, as shown in Ref.~\cite{0295-5075-100-3-38005}, fat--tails or
related stylized facts do not occur in this case. The RandomTraders
provide a neutral background for the trading activities of the
$N_\textrm{Ising}=144$ IsingTraders. The IsingTraders' parameters are
set to $J=1.0$, $\alpha = 4.0$ and $\beta=1.45$ in accordance with
Ref.~\cite{doi:10.1142/S0129183101001845}. The mean value
$\mu_\textrm{wt,Ising}$ of the waiting time
distribution~(\ref{eq:waitingtime}) for the IsingTraders only scales
with the number of IsingTraders,
\begin{equation}
  \mu_\textrm{wt,Ising}=cN_{\text{Ising}} \ .
\end{equation}
A good value for the sweep probability was found to be
$q_{\text{sweep}}=0.001$. To this end we looked at the price
stability resembled in the ratio of limit order placing RandomTraders
and market order placing IsingTraders. Importantly, the simulations
only very weakly depend on the exact value of $q_{\text{sweep}}$.

A crucial quantity is the average ratio $Q$ of IsingTraders who can
place their orders before the lattice is updated for the first time.
It is easily seen to be
\begin{align}
 Q&=\sum\limits_{n=1}^\infty q_{\text{sweep}}(1-q_{\text{sweep}})^{n-1}
 \left( 1- \mathrm{e}^{(n-1)/\mu_\textrm{wt,Ising}} \right)  \\
 &\approx 0.78 \ .
\end{align}
Knowledge of $Q$ gives a grip on the interplay between lattice and trading
dynamics.  This is so, because only if the lattice is updated slowly
enough the simulation ``sees'' something of the lattice dynamics.
\begin{figure*}[htbp]
  \begin{center}
	\subfloat{
    \includegraphics[width=\subimage\textwidth]{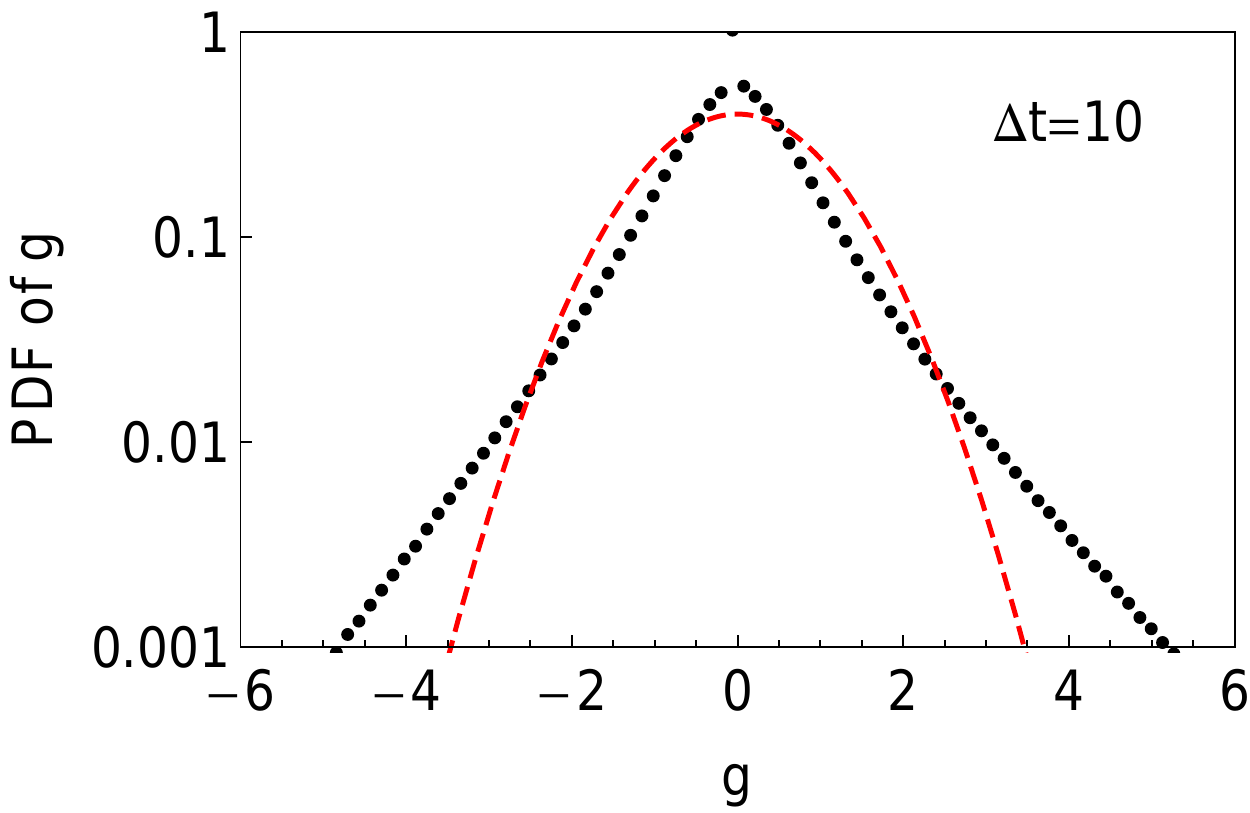}
\label{fig:results:rt_pdf:a}
}
	\subfloat{
    \includegraphics[width=\subimage\textwidth]{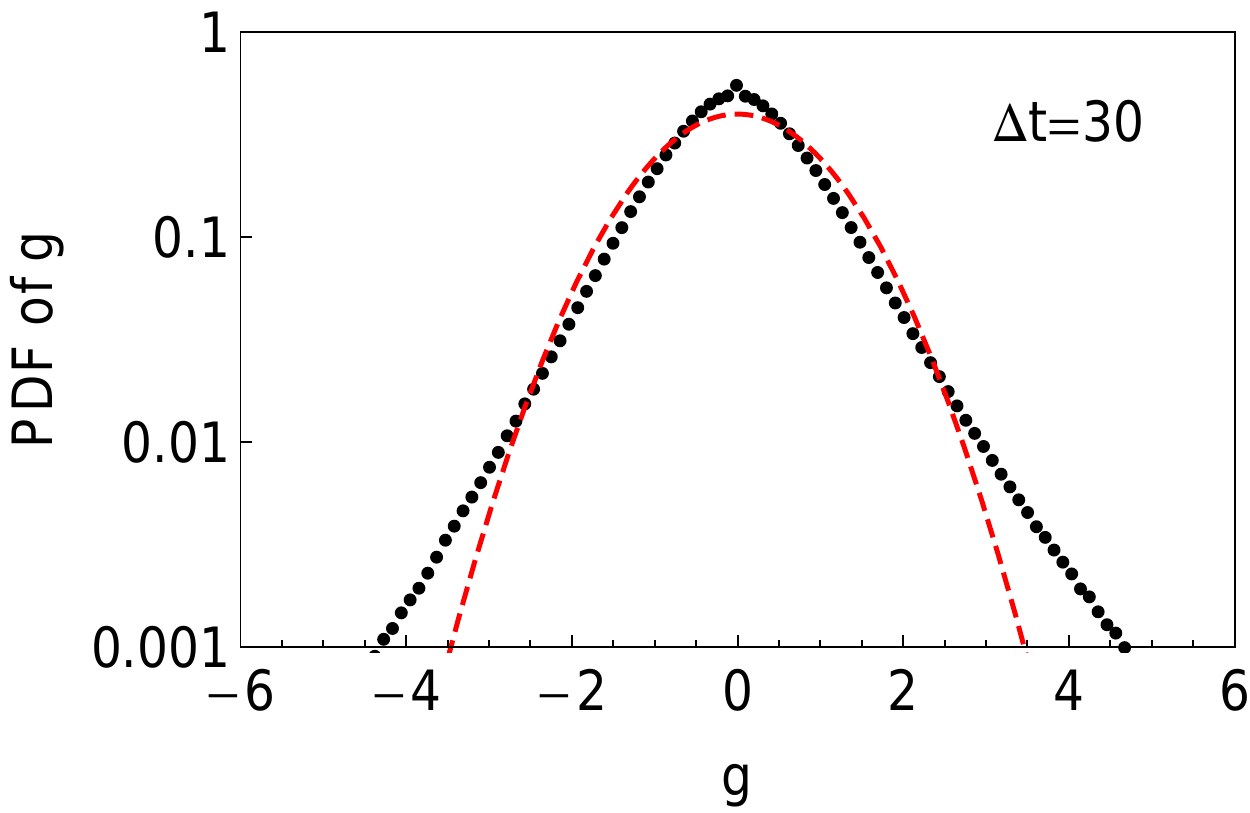}
\label{fig:results:rt_pdf:b}
}

	\subfloat{
    \includegraphics[width=\subimage\textwidth]{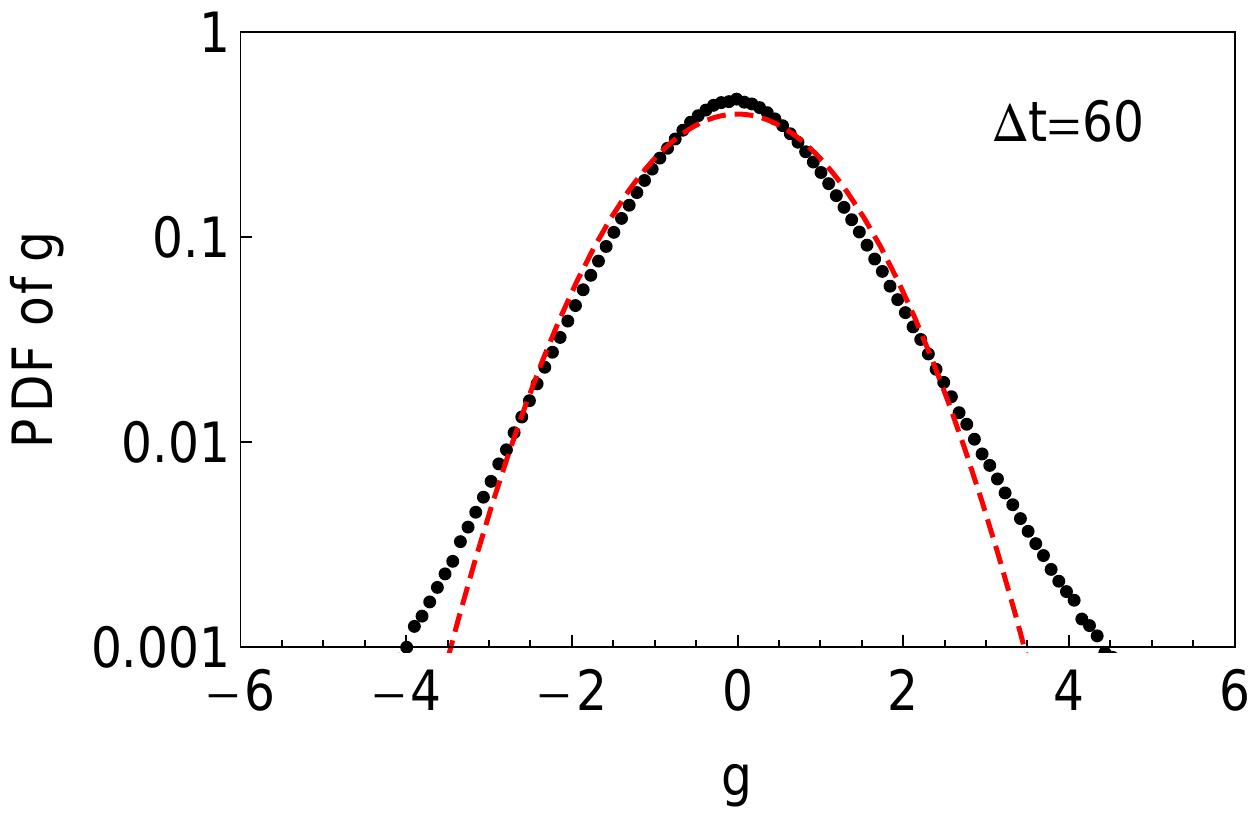}
\label{fig:results:rt_pdf:c}
}
	\subfloat{
    \includegraphics[width=\subimage\textwidth]{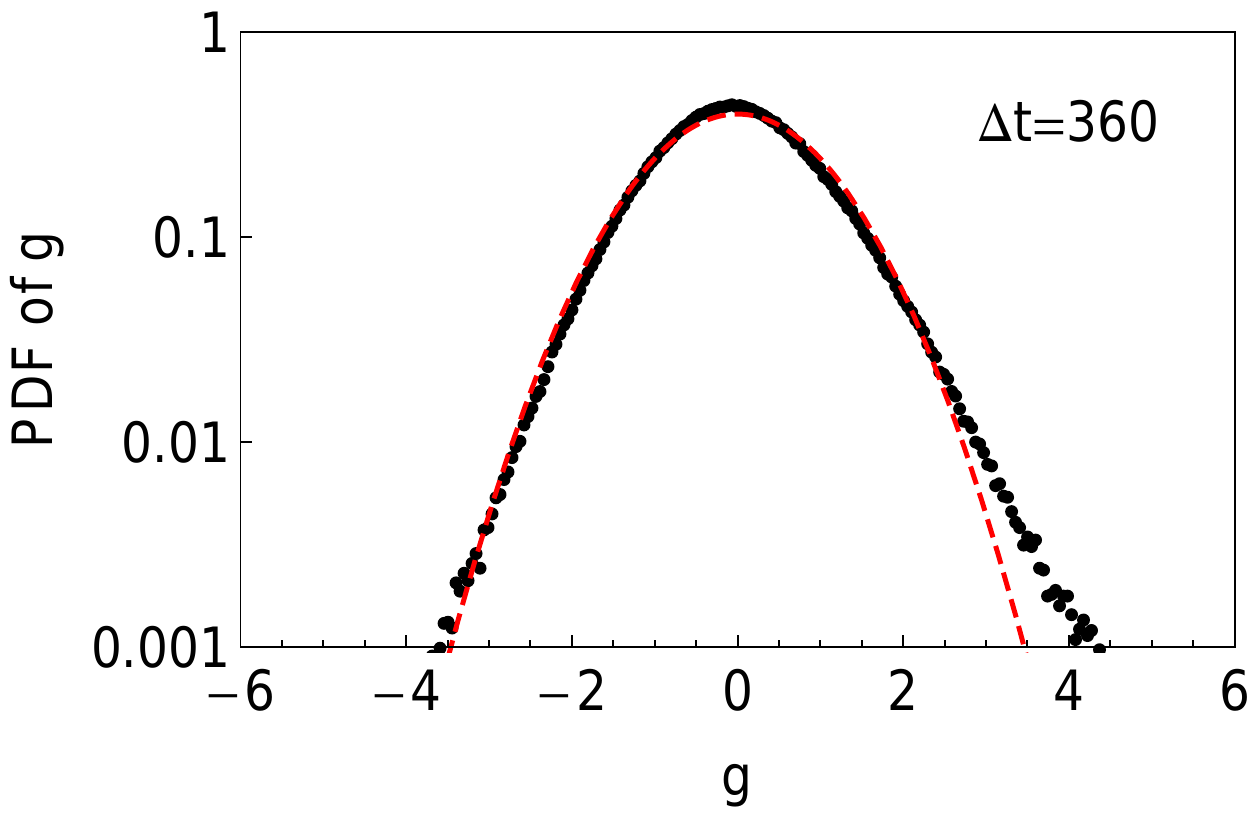}
\label{fig:results:rt_pdf:d}
}

	\subfloat{
    \includegraphics[width=\subimage\textwidth]{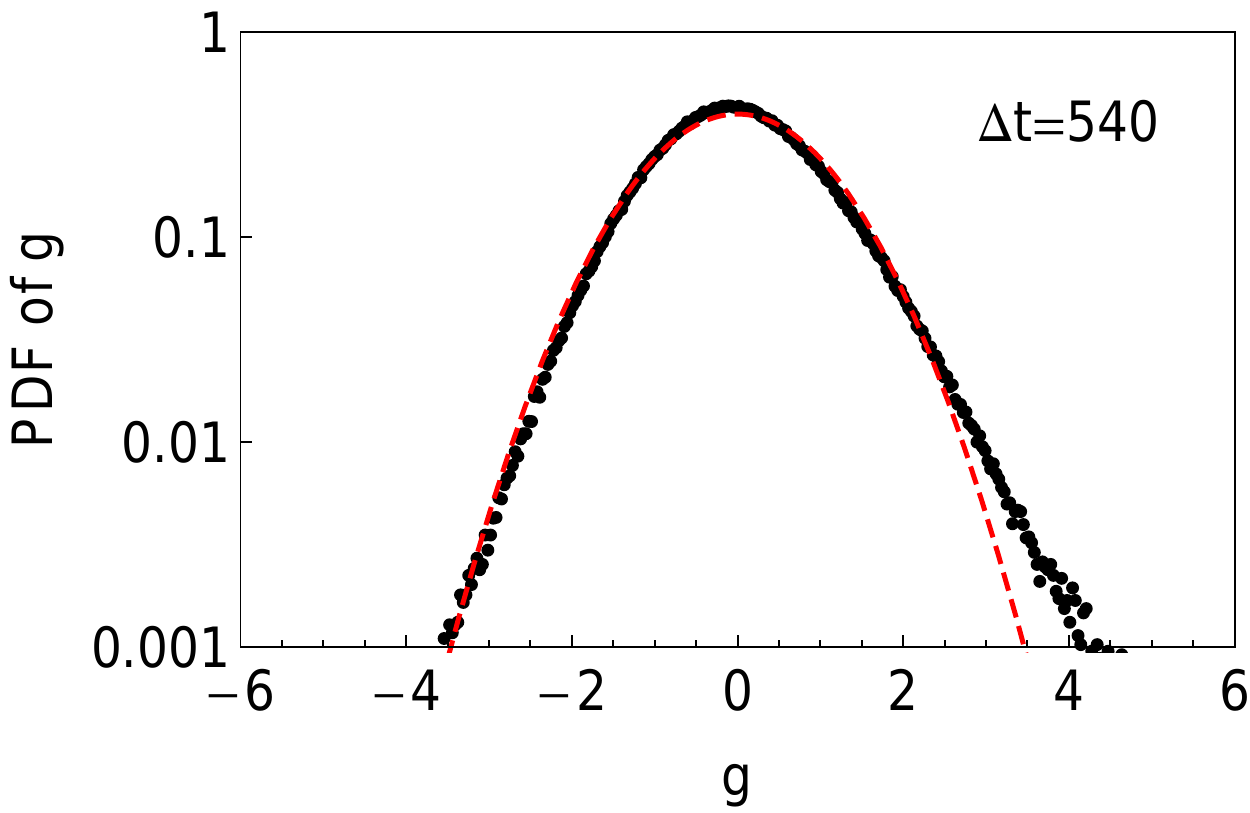}
\label{fig:results:rt_pdf:e}
}
	\subfloat{
    \includegraphics[width=\subimage\textwidth]{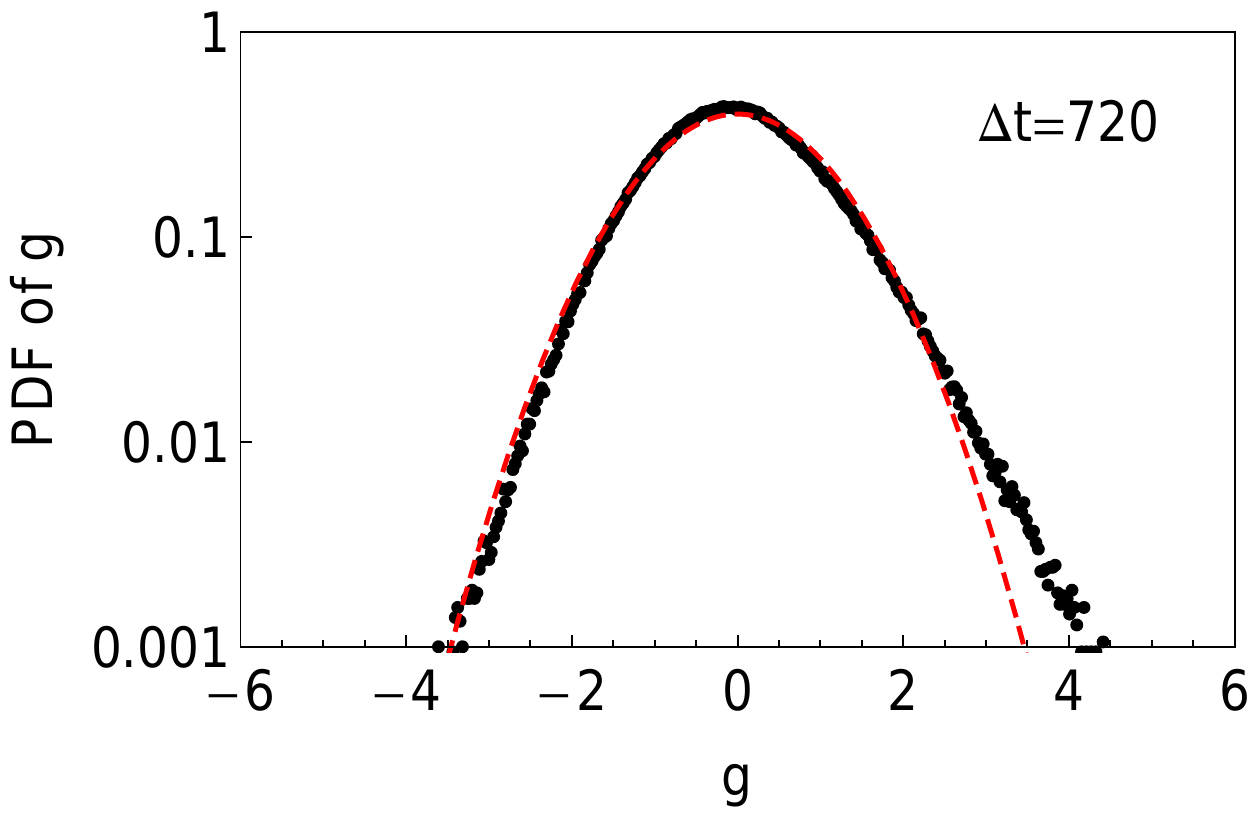}
\label{fig:results:rt_pdf:f}
}
  \end{center}
  \caption{Distributions of the normalized returns $g$ on a
    logarithmic scale for six return interval $\Delta t=10$, 30, 60, 360, 540 and 720 time steps. 
    A normal distribution is shown as dashed line.}
 \label{fig:results:rt_pdf}
\end{figure*}
Altogether, these choices yield on average a trading frequency of 5.4
trades per time step. We ran 1000 simulations with $T=5\cdot 10^5$
time steps. We checked that all of them were stable far beyond the
chosen time scale.

From the traded prices $p(t)$, we calculate the returns
\begin{equation}
 r(t)=\frac{p(t+\Delta t)-p(t)}{p(t)}
\end{equation}
with return interval $\Delta t$ as well as the the standard deviation
\begin{equation}
\sigma=\sqrt{\langle r(t)^2\rangle_{T}-\langle r(t)\rangle_{T}^{2}} \ ,
\end{equation}
where the angular brackets indicate the sample mean over the entire
trading time $T$. 
\begin{figure*}[htbp]
  \begin{center}
	\subfloat{
    \includegraphics[width=\subimage\textwidth]{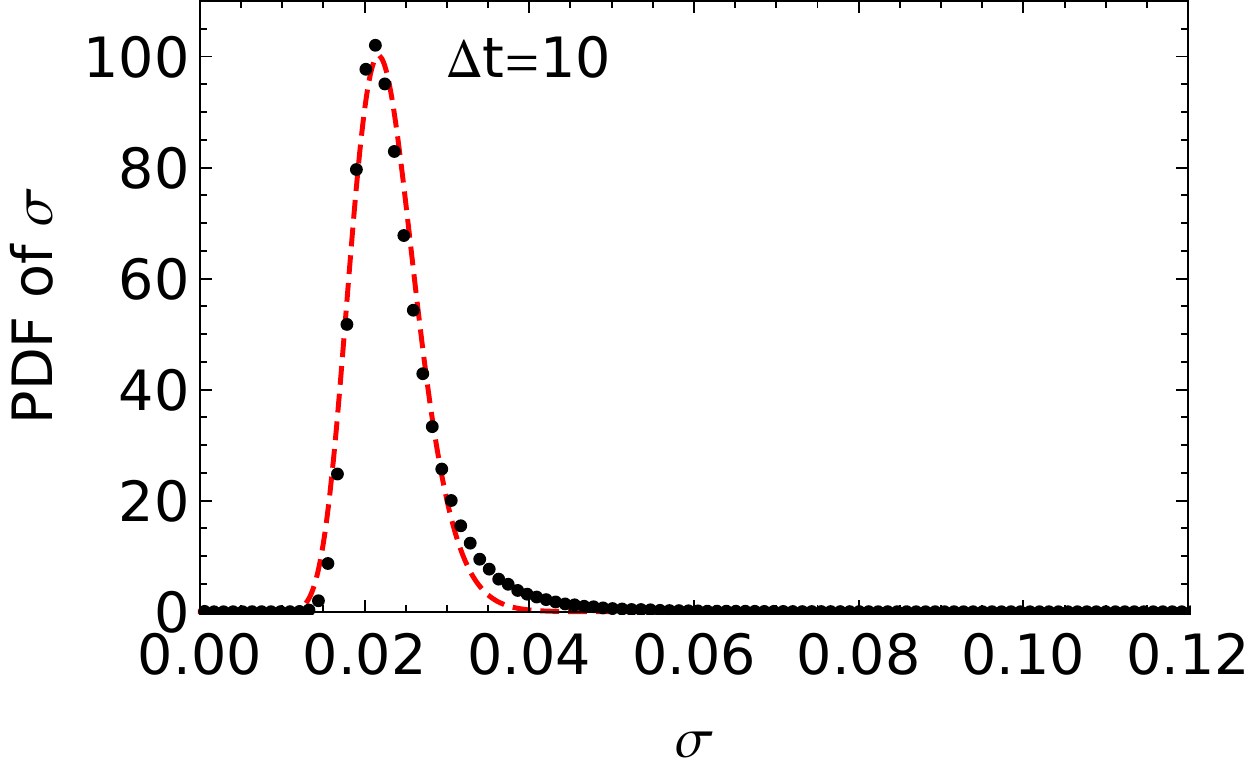}
\label{fig:results:vol_pdf:a}
}
	\subfloat{
    \includegraphics[width=\subimage\textwidth]{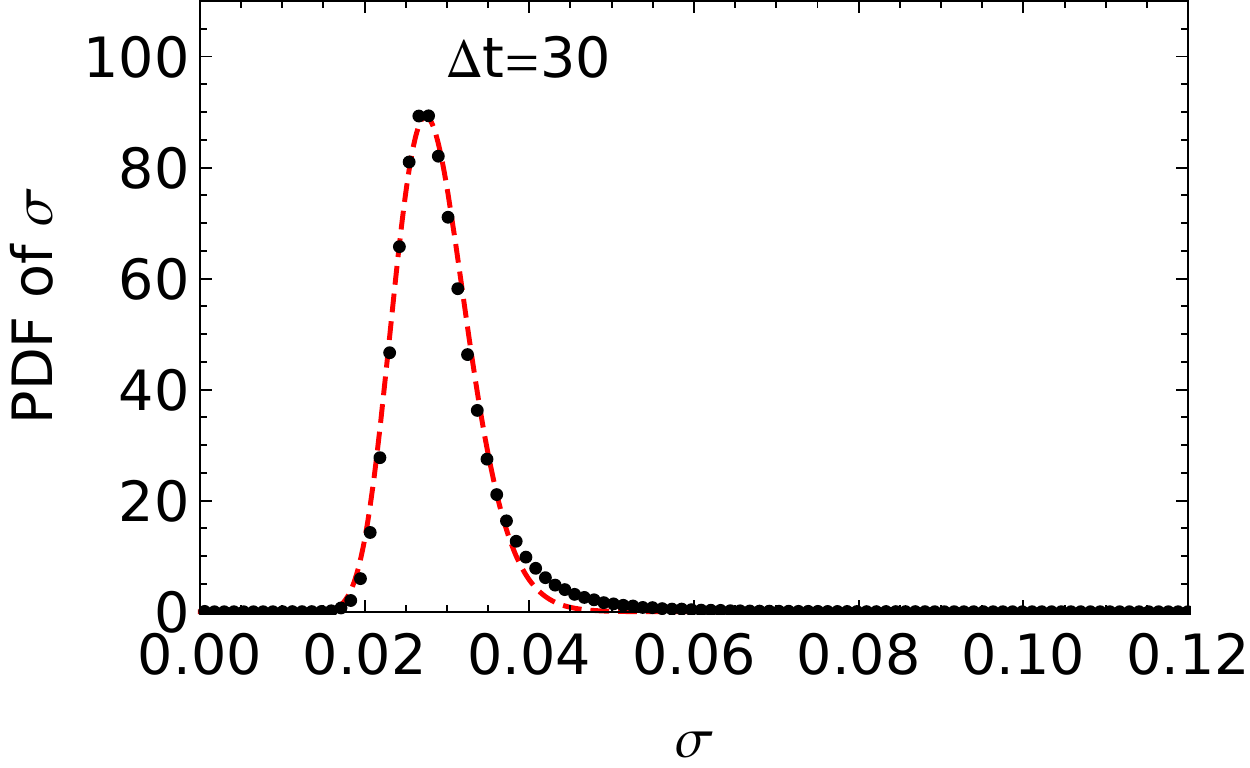}
\label{fig:results:vol_pdf:b}
}

	\subfloat{
    \includegraphics[width=\subimage\textwidth]{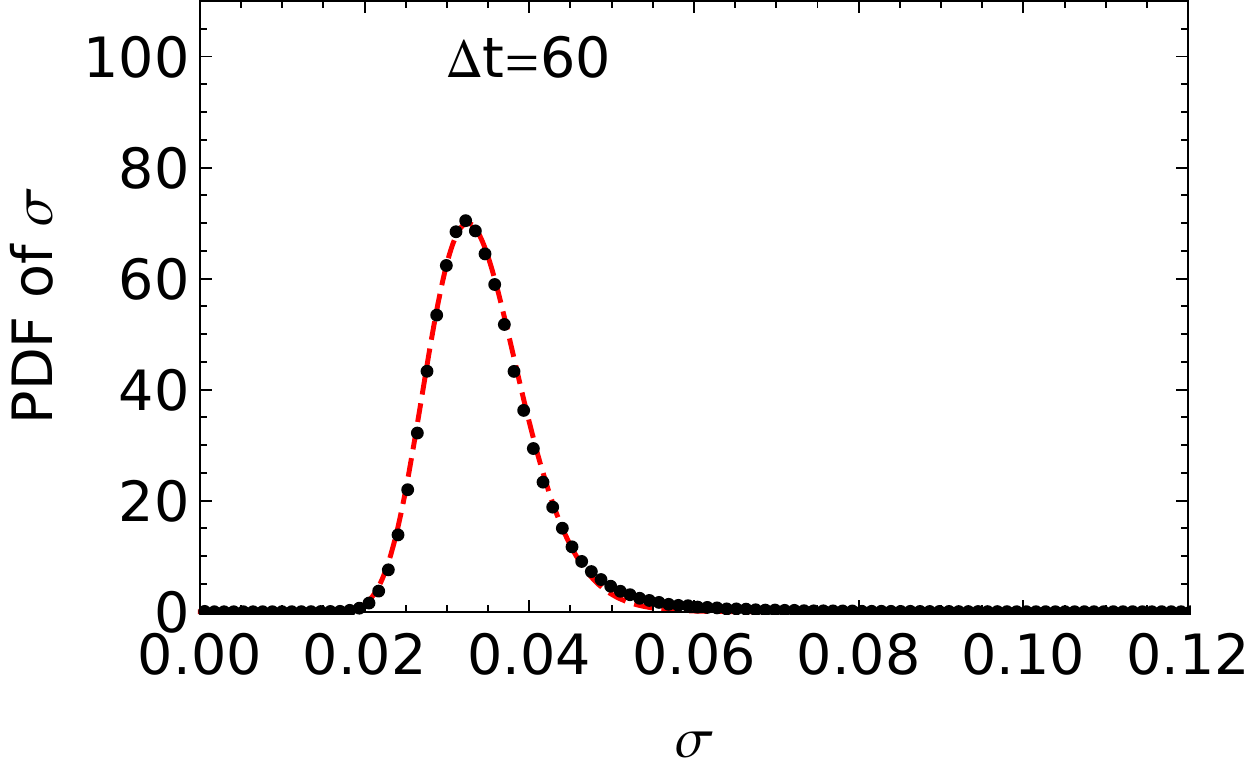}
\label{fig:results:vol_pdf:c}
}
	\subfloat{
    \includegraphics[width=\subimage\textwidth]{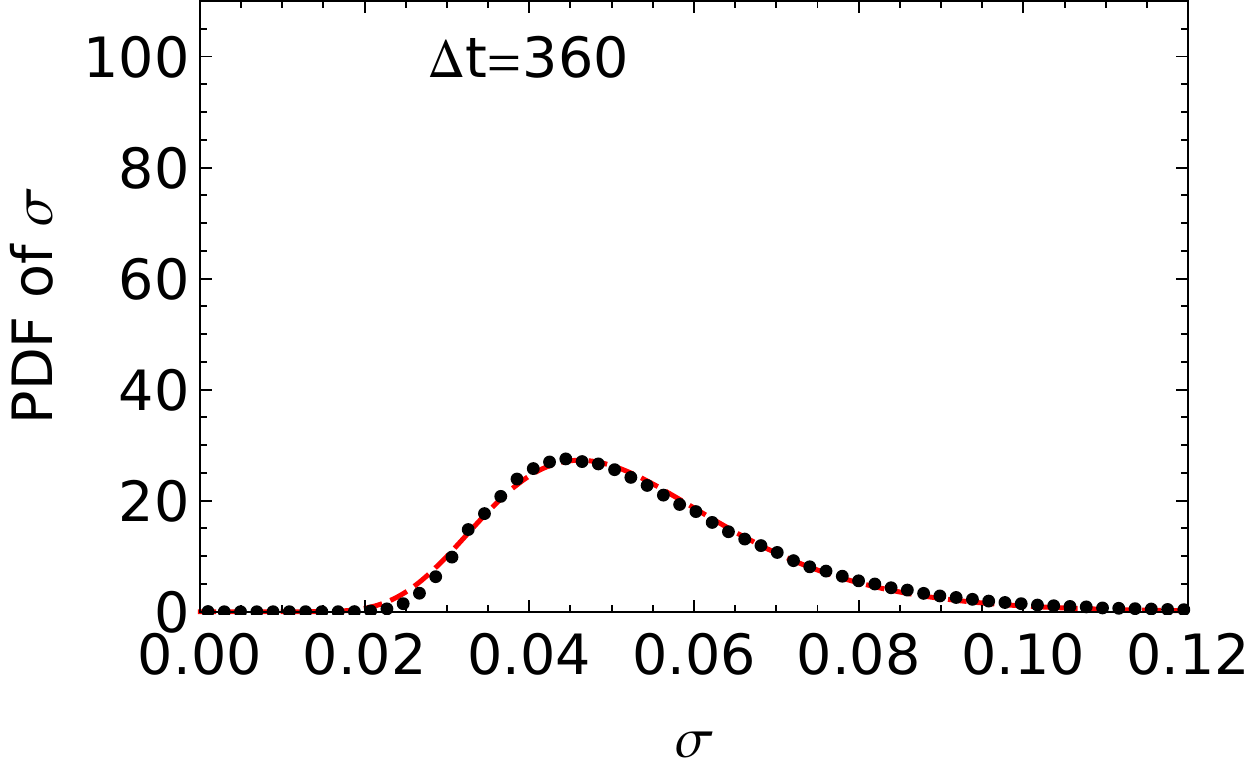}
\label{fig:results:vol_pdf:d}
}

	\subfloat{
    \includegraphics[width=\subimage\textwidth]{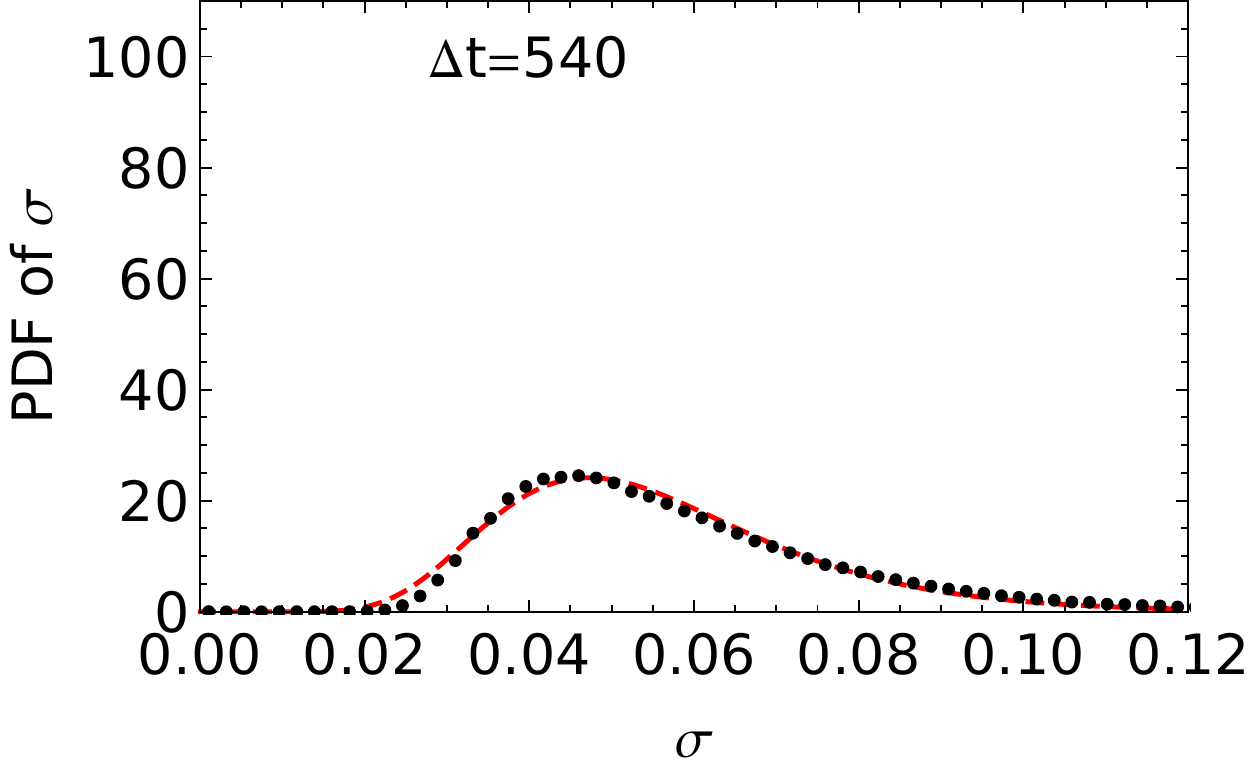}
\label{fig:results:vol_pdf:e}
}
	\subfloat{
    \includegraphics[width=\subimage\textwidth]{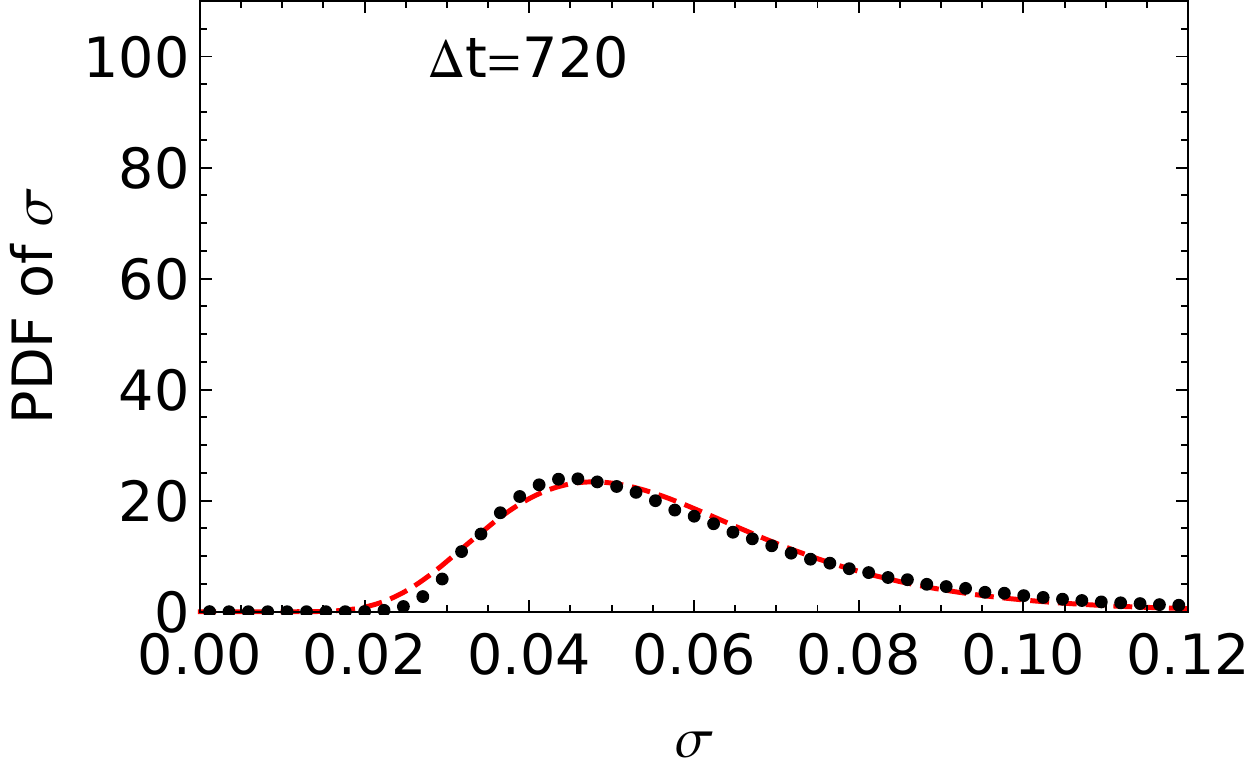}
\label{fig:results:vol_pdf:f}
}
  \end{center}
  \caption{Volatility distributions for six different return
    intervals $\Delta t$. A log--normal distribution is shown as dashed
    line.}
 \label{fig:results:vol_pdf}
\end{figure*}

To study the behavior of the return distribution, we normalize the
returns to zero mean and unit standard deviation,
\begin{equation}
 g(t)=\frac{r(t)-\langle r(t)\rangle_{T}}
                 {\sigma} \ .
\end{equation}
In Fig.~\ref{fig:results:rt_pdf}, we compare the distributions of the
normalized returns $g(t)$ for six different return intervals $\Delta
t$ to a normal distribution. We observe heavy tails which are the
lower the larger the return intervals $\Delta t$.

To also investigate the volatility distributions, we calculate time
dependent volatilities by moving a window of $1000$ time steps through
the data.  The resulting distributions for different return intervals
are displayed in Fig.~\ref{fig:results:vol_pdf}. They agree well with
a log--normal distribution, which is consistent with empirically found
volatility distributions.~\cite{Micciche2002756}.

\begin{figure*}[htbp]
  \begin{center}
	\subfloat{
    \includegraphics[width=\subimage\textwidth]{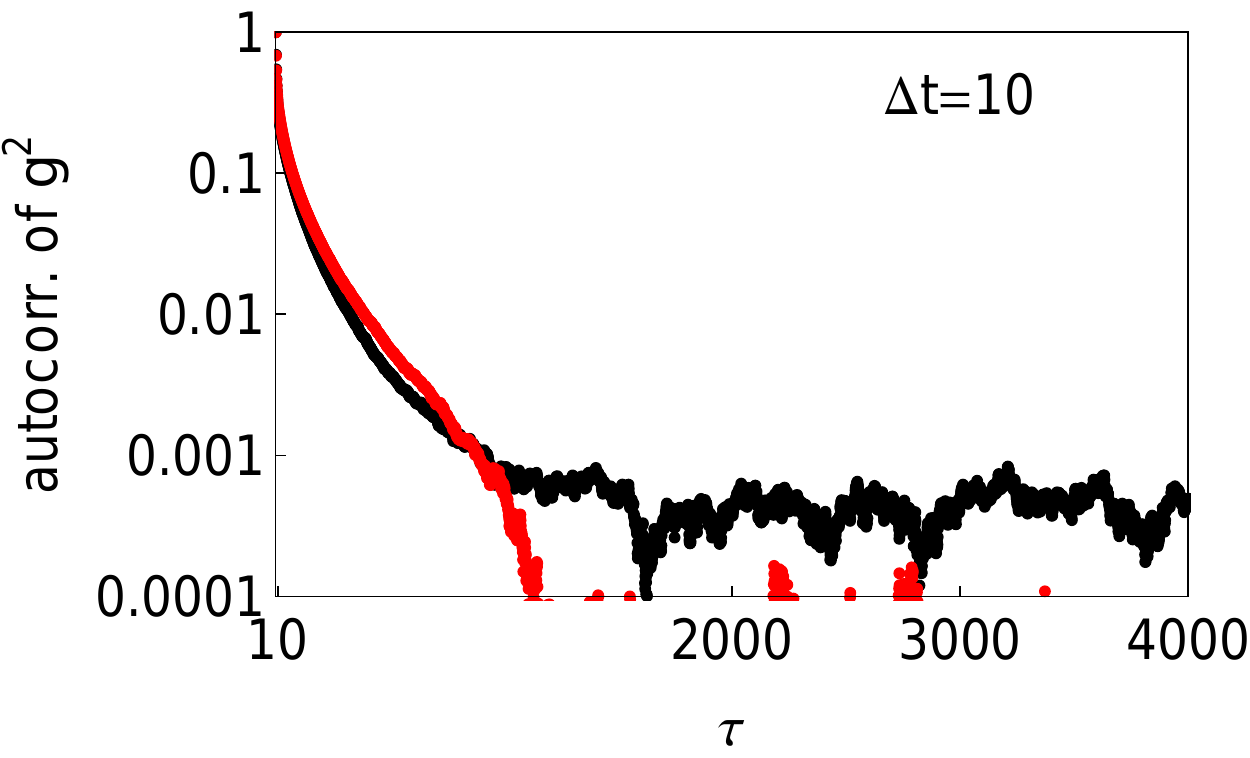}
\label{fig:results:auto_pdf:a}
}
	\subfloat{
    \includegraphics[width=\subimage\textwidth]{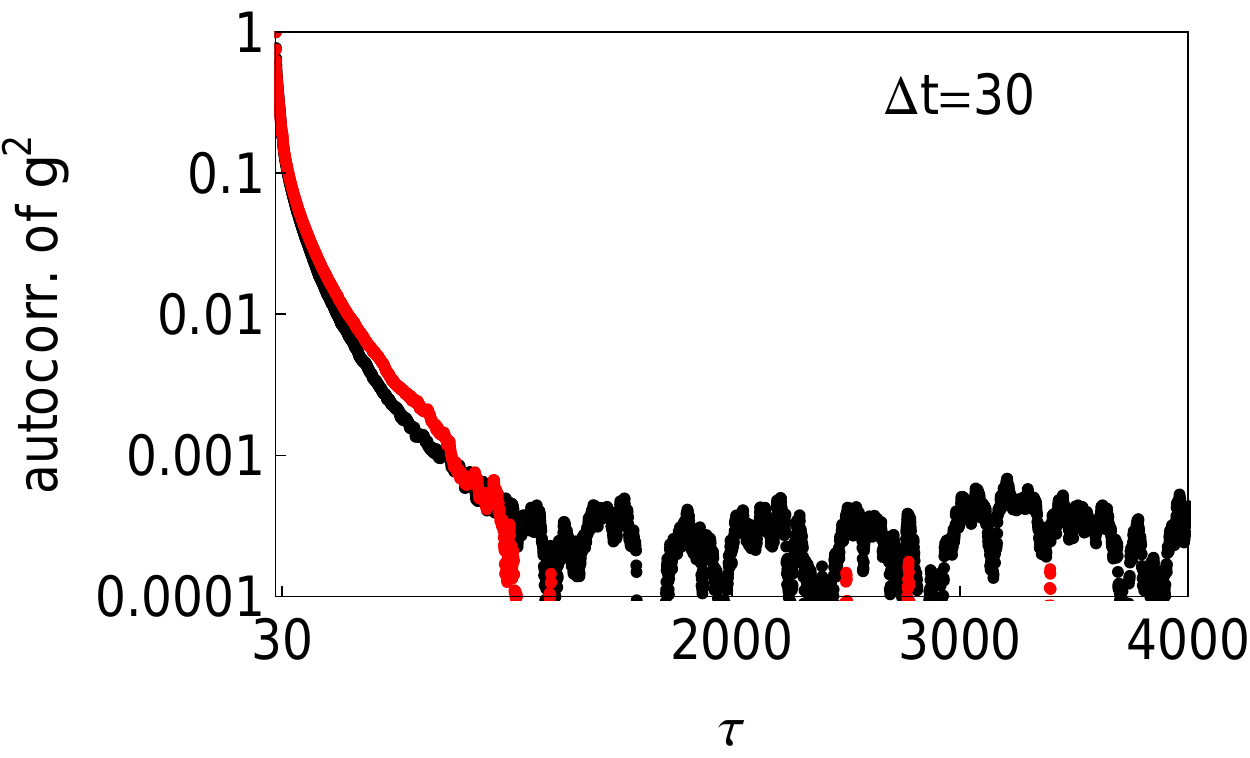}
\label{fig:results:auto_pdf:b}
}

	\subfloat{
    \includegraphics[width=\subimage\textwidth]{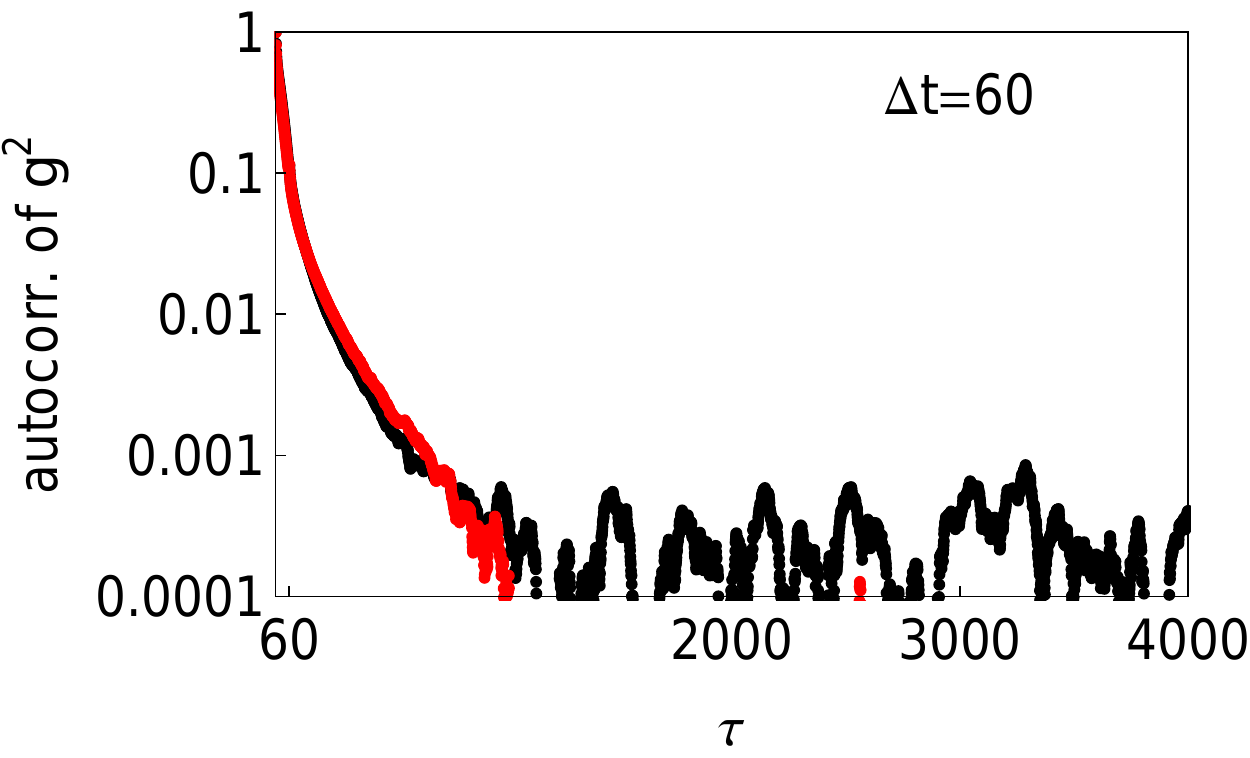}
\label{fig:results:auto_pdf:c}
}
	\subfloat{
    \includegraphics[width=\subimage\textwidth]{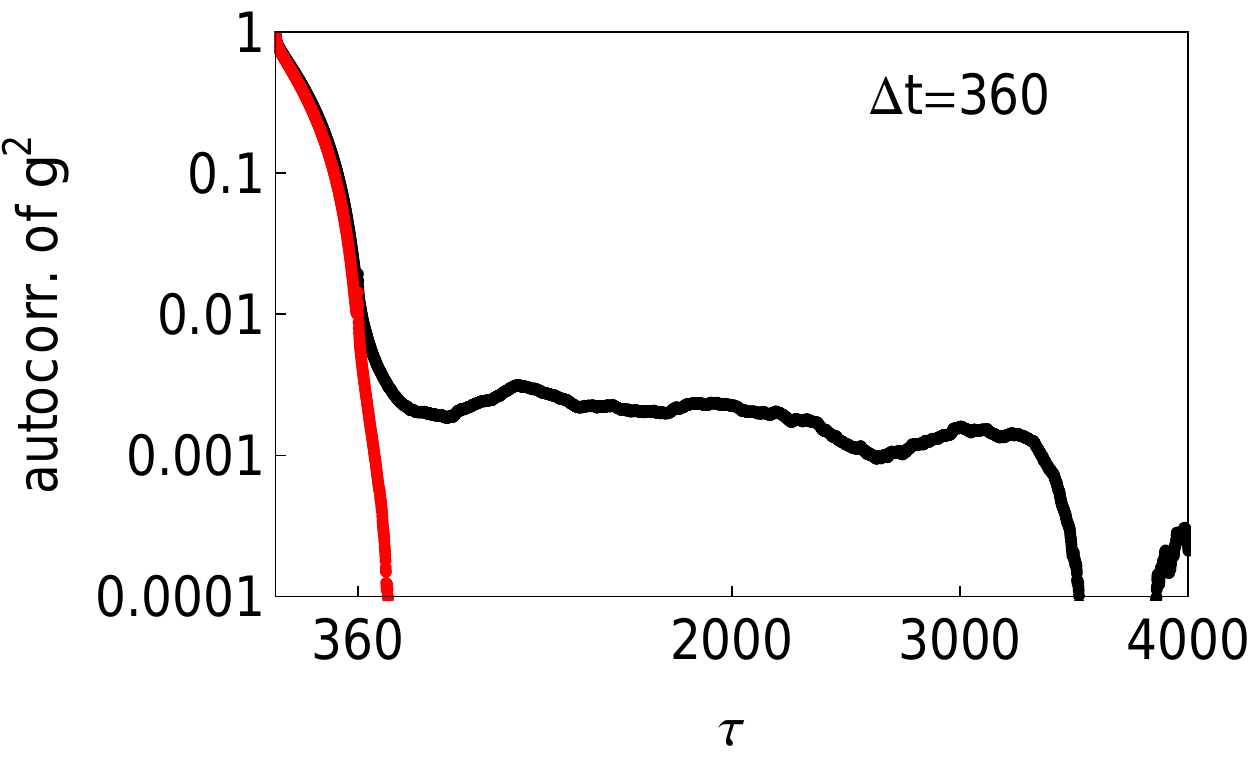}
\label{fig:results:auto_pdf:d}
}

	\subfloat{
    \includegraphics[width=\subimage\textwidth]{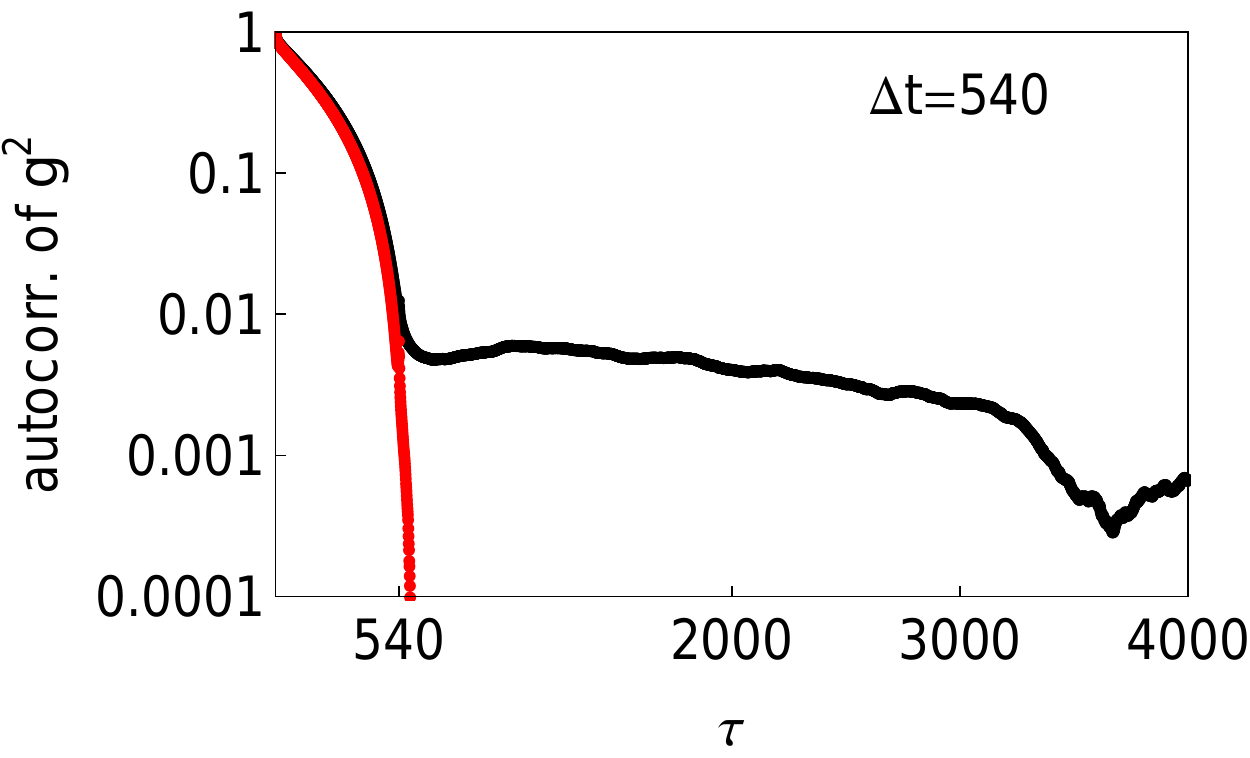}
\label{fig:results:auto_pdf:e}
}
	\subfloat{
    \includegraphics[width=\subimage\textwidth]{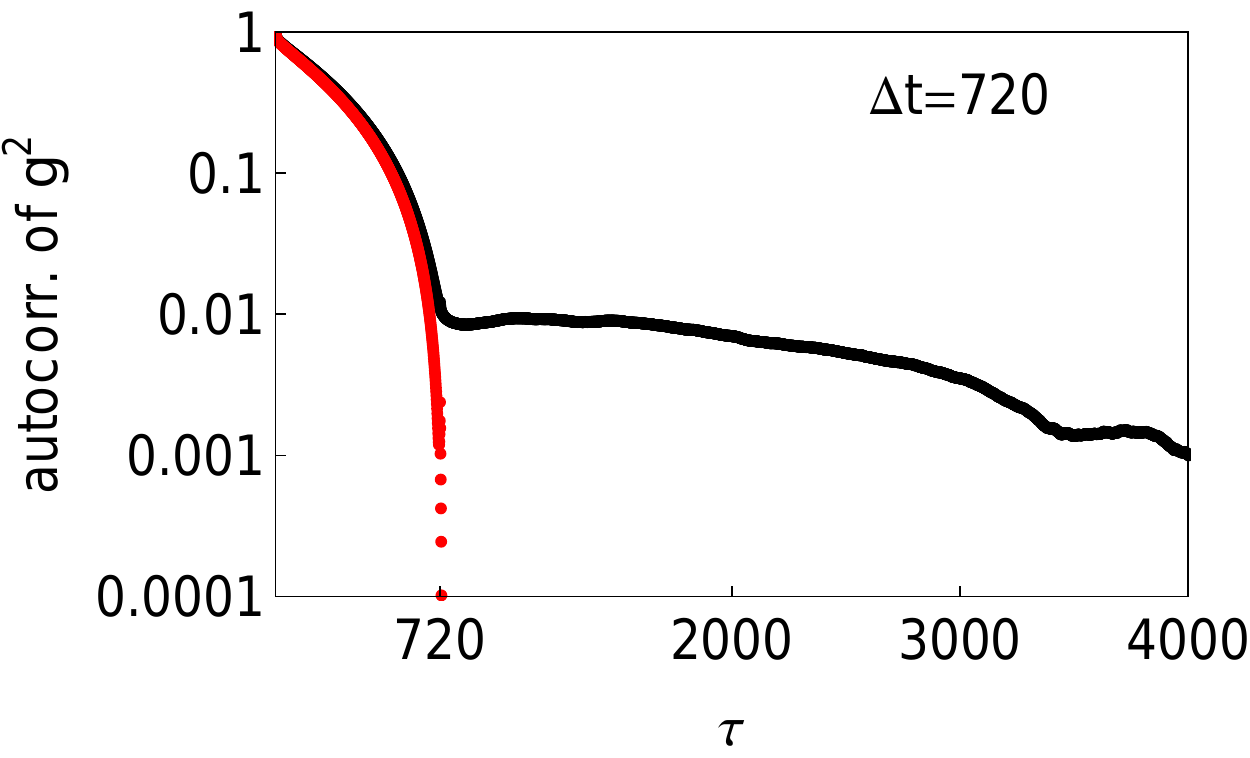}
\label{fig:results:auto_pdf:f}
}
  \end{center}
  \caption{Autocorrelation function of the squared returns $g^2$ for
    six different return intervals $\Delta t$ versus time lag $\tau$, for LiquidityTakers as red and for IsingTraders as black
    lines.}
 \label{fig:results:auto_pdf}
\end{figure*}

What is the effect of the IsingTrader's strategy? --- In
Fig.~\ref{fig:results:auto_pdf}, We compare his time autocorrelation
\begin{equation}
\textrm{acf}(\tau) = \langle g^2(t) g^2(t+\tau)\rangle_{T}
\end{equation}
for squared returns as function of a time lag $\tau$ to that of
the LiquidityTaker. Interestingly, the IsingTrader creates long--range
autocorrelation independently of the chosen return interval. A similar
behavior has indeed been observed in empirical
data~\cite{doi:10.1080/713665670}. This behavior can be traced back to
the autocorrelation of the squared magnetization changes as observed
in Ref.~\cite{doi:10.1142/S0129183101001845}.  However, in the case of
small return intervals, we also observe a non zero autocorrelation for
lags greater than the return interval for the LiquidityTaker. This is
a result of the rate of trades compared to the return interval. If the
trade frequency is very low compared to the return interval, the price
does not change and we obtain many successive zero returns. This can
be seen in Fig.~\ref{fig:results:rt_pdf} for $\Delta t=10$ and $\Delta t=30$. A persistent autocorrelation of the
squared returns results. We therefore find similar behavior for the
LiqudityTaker and the IsingTrader at relatively small lags. It is less
pronounced and eventually disappears when the return interval
increases.

\section{Conclusion}
\label{sec5}

Models for financial markets typically consist of two parts: decision
making and price formation. Due to the \textit{a priori} limited
information about the individual trader, statistical concepts have to
be invoked to model the decision making. The challenge is then to
capture salient features of the highly complex dynamics by stochastic
ingredients. On the other hand, the price formation is, in real stock
exchanges, a microscopically well defined and deterministic
process. The reason why the price formation part of most models relies
on equilibrium pricing and does not involve the order book is probably
twofold: First, the concept of equilibrium pricing is deeply rooted in
the economics literature.  Second, even though the order book dynamics
is microscopically well defined, it is not possible to directly map it
on a simple schematic equation.

Hence, we find it worthwhile to critically examine the price formation
part of stock market models by putting their specific decision making
part in a realistic order book setting. In a previous
study~\cite{Wagner2014347}, we carried out this program for a model whose
decision making part is much simpler than in the
Bornholdt--Kaizoji--Fujiwara model that we investigated in the present
study. The price formation part of the latter explicitly employs the concept of a
``fundamental'' price which is as questionable as the closely related
concept of a ``fair'' price in the efficient market model. To model
the dynamics of a system, \textit{i.e.}, the stock market in the
present case, only quantities should enter which have some empirical
justification. The ``fundamental'' price, however, is an external
criterion that cannot, not even indirectly, be measured.

We implemented the decision making part of the
Bornholdt--Kaizoji--Fujiwara model in a minimalistic agent--based
model and performed numerical simulations. They yield realistic
stylized facts, including non--trivial features such as long--range
temporal autocorrelations. This implies that the equilibrium pricing
mechanism is fully obsolete.  We thus conclude that the decision
making part of the Bornholdt--Kaizoji--Fujiwara model grasps essential
features of the market dynamics in a reliable manner.

Furthermore, we may draw a second conclusion. In an indirect way, our
study shows that the equilibrium pricing mechanism can be used to
coarsely and effectively mimick the order book dynamics. Put
differently, the highly complex order book dynamics generates
\textit{upon average} a schematic rule that seems to be largely
equivalent to an equilibrium pricing mechanism. Contrary to the common
assumption in economics, there is no need whatsoever to require that
the \textit{individual} trades result from such an equilibrium pricing
mechanism.  Admittedly, we can base these statements only on the
present case study, but we are tempted to believe that they are more
general. Nevertheless, it is not necessary to evoke the sometimes
almost ideological reasoning behind the concept of equilibrium pricing.
Abandoning such a strict and schematic view also helps to understand
turbulent market situations in which the assumption of equilibrium
pricing is even less plausible than during quiet times. Equilibrium
pricing should at best only be seen as an averaged result of the true
dynamics: Everything is in the order book.

\section*{Acknowledgments}

We thank S.~Bornholdt, T.~Kaizoji and Y.~Fujiwara for making their
preliminary studies on the spin model available to us.

\end{document}